\documentclass{elsart}

\usepackage{natbib}
\usepackage{graphicx}
\usepackage{amssymb}

\newcommand{\rbrkt}[1]{\left(#1\right)}

\begin{document}

\begin{frontmatter}

\title{Size Effects and Dislocation Patterning in \\Two-Dimensional Bending}

\author{N. Scott Weingarten}

\address{Physics Department, The Catholic University of America, Washington, DC}

\author{Robin L. B. Selinger}

\address{Chemical Physics Interdisciplinary Program and Liquid Crystal
Institute, Kent State University, Kent, OH \\Physics Department, The
Catholic University of America, Washington, DC}


\begin{abstract}
We perform atomistic Monte Carlo simulations of bending a Lennard-Jones single crystal in two dimensions. Dislocations nucleate only at the free surface as there are no sources in the interior of the sample.  When dislocations reach sufficient density, they spontaneously coalesce to nucleate grain boundaries, and the resulting microstructure depends strongly on the initial crystal orientation of the sample. In initial yield, we find a reverse size effect, in which larger samples show a higher scaled bending moment than smaller samples for a given strain and strain rate. This effect is associated with source-limited plasticity and high strain rate relative to dislocation mobility, and the size effect in initial yield disappears when we scale the data to account for strain rate effects. Once dislocations coalesce to form grain boundaries, the size effect reverses and we find that smaller crystals support a higher scaled bending moment than larger crystals. This finding is in qualitative agreement with experimental results. Finally, we observe an instability at the compressed crystal surface that suggests a novel mechanism for the formation of a hillock structure. The hillock is formed when a high angle grain boundary, after absorbing additional dislocations, becomes unstable and folds to form a new crystal grain that protrudes from the free surface.
\end{abstract}

\end{frontmatter}

\section{Introduction}
One unsolved mystery in understanding the mechanical response of crystalline solids is the observation that sample size alters mechanical response in plastic deformation.  Such size effects have been observed experimentally in crystals deformed under torsion~\citep{fleck}, bending~\citep{stolken-evans, soboyejo}, and nanoindentation~\citep{mcelhaney,suresh}, geometries which all involve strain gradients.  Size effects have also recently been observed in geometries without strain gradients~\citep{espinosa}.  Attempts to fit size effects into a plasticity model have included the development of higher order strain gradient theories, with characteristic length parameters at the micron scale ~\citep{fleck-hutch, gurtin, gao, deborst} whose physical meaning remains a subject of ongoing research.

In the present work we address the question: do size effects occur in the simplest possible system, a single crystal in two dimensions? While not closely relevant to the properties of real materials, atomistic simulations in two dimensions can reveal fundamental mechanisms in dislocation dynamics and interactions. Like the bubble-raft experiments of~\citet{bragg}, they offer easy visualization and analysis, with a restricted population of defect types, that is, edge dislocations on three slip systems. Bubble raft experiments continue to provide new insights into the mechanics of materials~\citep{gouldstone}.

Dislocation patterning is another process whose mechanisms are not entirely understood and which strongly affects a material's mechanical response~\citep{hughes}.  Simulation studies of such patterning have been performed extensively using dislocation dynamics (DD) techniques at the mesoscale level and via reaction-diffusion models~\citep{groma, aifantis86, deshpande03, devincre}. Atomistic simulations can validate these mesoscale models and identify mechanisms that control the nucleation and evolution of dislocation patterns. Atomistic simulations also provide the opportunity to look at temperature effects.

Both size effects and dislocation patterning involve the emergence of characteristic length scales, though patterning can also involve fractal scaling~\citep{zaiser}.  Simulation studies of size effects in plastic deformation have been performed at the atomistic scale~\citep{baskes}, and on the mesoscale~\citep{needleman, shu}. ~\citet{deshpande05} recently carried out a beautiful DD investigation of size effects in a two-dimensional single crystal using a technique that allows unconstrained rotation of the tensile axis. The geometry of their simulations differs from the present work in several respects, but the most important difference is that they include a random distribution of dislocation sources and obstacles in the interior of the crystal, while in the present work, the sample contains no pre-existing sources or obstacles. Any imposed density of sources already defines a characteristic length scale, namely, the average spacing between sources, so it is perhaps less surprising that such a simulation will show size effects. Horstemeyer et. al. (2001) postulated the emergence of a characteristic length scale related to the volume to surface area ratio, but subsequent experiments by~\citet{greer} on compression of gold pillars have shown contradictory behavior.

More recently,~\cite{wu} have developed a novel experimental approach to bending nanowires. While they observe a clear size effect in initial yield for untreated polycrystalline gold nanowires, they observed no clear size effect for annealed gold nanowires, where initial density of dislocations and grain boundaries is greatly reduced. As will be discussed, our simulations suggest that the initial yield strength of single crystals is independent of sample size, once strain rate effects are taken into account, in qualitative agreement with Wu's experimental data. In our simulations we observe size effects in mechanical response only after dislocation density increases enough to initiate the formation of grain boundaries.

\section{Simulation methods}
\begin{figure}[t]
\centering
\includegraphics[width=4in]{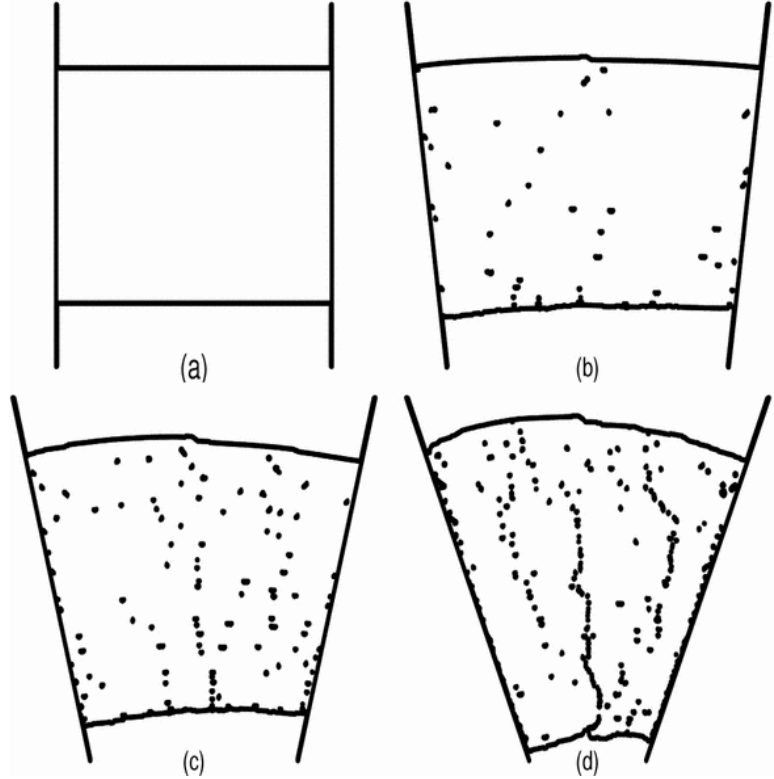}
\caption{Two dimensional Lennard-Jones crystal confined between rigid
walls, at bending strain of (a) 0, (b) 0.15, (c) 0.3, and (d) 0.5; sample consists of $73,879$ particles, but only atoms with other than six nearest neighbors are shown.}
\label{fig:bendalgorithm}
\end{figure}

To gain insight into size effects and patterning in plastic deformation, we perform atomistic Monte Carlo simulations of bending a single crystal in two dimensions.  The model system is a triangular lattice of atoms interacting via a Lennard-Jones pair potential
\begin{equation}
V_{ij}=-4\epsilon\rbrkt{\rbrkt{\frac{\sigma}{r_{ij}}}^{6}-\rbrkt{\frac{\sigma}{r_{ij}}}^{12}}
\end{equation}
with parameters $\sigma=\epsilon=1$, at temperature $k_b T = 0.1$. The two-dimensional Lennard-Jones crystal has been used previously for both Monte Carlo and molecular dynamics studies of dislocation nucleation and ideal strength as a function of temperature~\citep{selinger1, selinger2}. The sample is confined between rigid walls of the same material that rotate inwards, as seen in Movie~1\footnote{Movie 1 available at http://www.lci.kent.edu/{$\sim$}robin/Movie1.avi}, with open boundary conditions above and below.  Fig.~\ref{fig:bendalgorithm} shows the configuration of the sample at four different strains. To visualize lattice defects, only atoms with fewer or more than six nearest neighbors are displayed. The top edge of the crystal is the neutral axis and the lower edge has the largest compression, with a strain gradient in between.  No region is under tension, reducing the chance of crack nucleation.  The applied strain increases linearly in discrete increments in which all particles are instantaneously displaced in the horizontal direction based on their position; a total bending strain of $50\%$ is achieved after $8,750$ strain increments.  Between strain increments, the system evolves via off-lattice Monte Carlo (MC) simulation using the Metropolis algorithm~\citep{allen-tildesley} with single particle moves.  While there is no real time scale, the effective strain rate is controlled by the number of MC steps (defined per particle) between strain increments.  Our longest simulations require $3.5$ million MC steps, with system sizes up to $360 \times 360$ particles.

Though the absence of a physical time scale is a drawback, the Monte Carlo method offers advantages compared to molecular dynamics (MD) for modelling plastic deformation at finite temperature. Because MC has no momentum, dislocation motion is always over-damped. Elastic waves cannot propagate so there is no need for damping zones at the boundaries to prevent reflections~\citep{holian-ravelo}.  In MC, each particle is coupled to a heat bath, while in MD the choice of thermostat can alter mechanical response, for instance changing whether a liquid is shear thickening or shear thinning~\citep{evans}.

To study size effects, simulation of bending was performed on samples ranging from $8,513$ to $123,879$ atoms.  The crystal was oriented with its three slip planes horizontal and at $\pm 60^\circ$.  The mechanical response of each sample is characterized by its bending moment as a function of strain.  The bending moment $M$ is defined as
\begin{equation}
M=\int{\sigma_{xx}(y)y{dy}}
\end{equation}
where $\sigma_{xx}(y)$ is the tensile stress profile through the sample.  This stress profile is determined by dividing the crystal into thin layers and calculating the average $\sigma_{xx}$ in each layer via the interatomic potential~\citep{allen-tildesley}.  To avoid artifacts associated with contact forces along the edges of the sample, we exclude the wall regions from the stress calculation.  The bending moment $M$ is then determined by numerical integration of the stress profile.

\section{Results}

\subsection{Size effects and the role of strain rate}
\begin{figure}[t]
\centering
\includegraphics[width=5in]{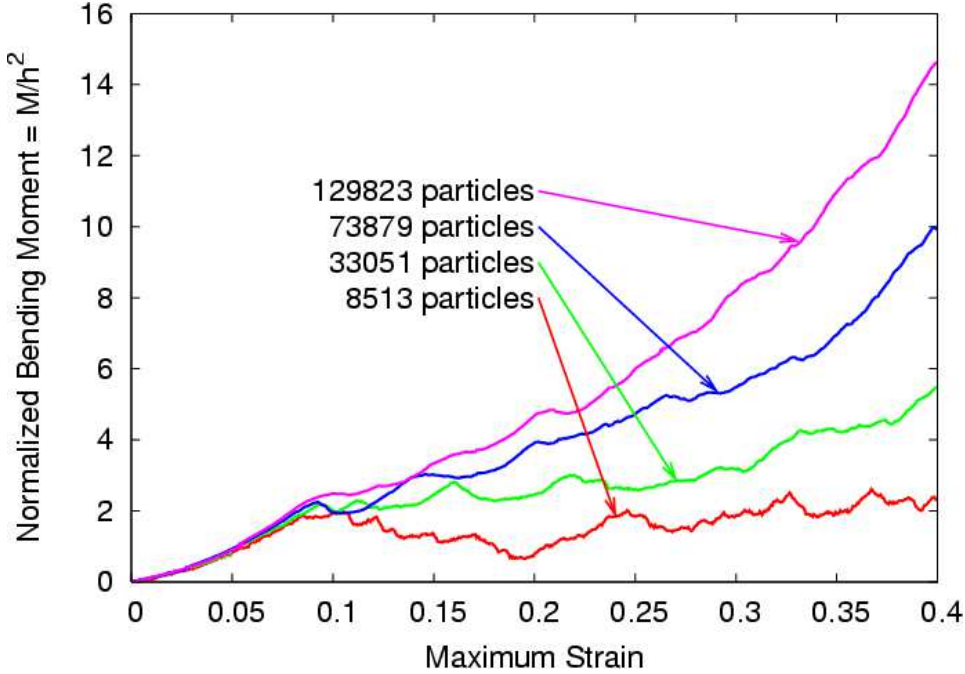}
\caption{(color) Normalized bending moment versus bending strain for four sample sizes, deformed at the same rate.  A reverse size effect is observed in initial yield.}
\label{fig:samerate}
\end{figure}

Fig.~\ref{fig:samerate} shows the normalized bending moment as a function of strain for four different crystal sizes, deformed at the same strain rate, e.g. with the same number of MC steps between strain increments. The bending moment is normalized by $h^2$ where $h$ is the sample height. After each sample is compressed to its elastic limit of about $5\%$ strain, dislocations begin to nucleate first at the two lower corners, where stress is most concentrated, as seen in Movies~1~and~2\footnote{Movie 2 available at http://www.lci.kent.edu/{$\sim$}robin/Movie2.avi}. Later, dislocations nucleate on the surfaces of the sample, but there are no dislocation sources in the bulk.

In our implementation of the bending algorithm, the particles comprising the two rigid confining walls are displaced only horizontally, even though the wall profile swings along a circular path. As a result, the spacing of particles along each wall lengthens gradually, creating misfit strain between the wall and the confined crystal. When the misfit reaches a high enough value, misfit dislocations nucleate along the confining walls.  Dislocations also pile up against the walls and sometimes dissociate. Stress concentration at steps on the unconfined free surfaces also gives rise to dislocation nucleation.  All of these processes contribute toward dislocation pattern formation.

\begin{figure}[t]
\centering
\includegraphics[width=5in]{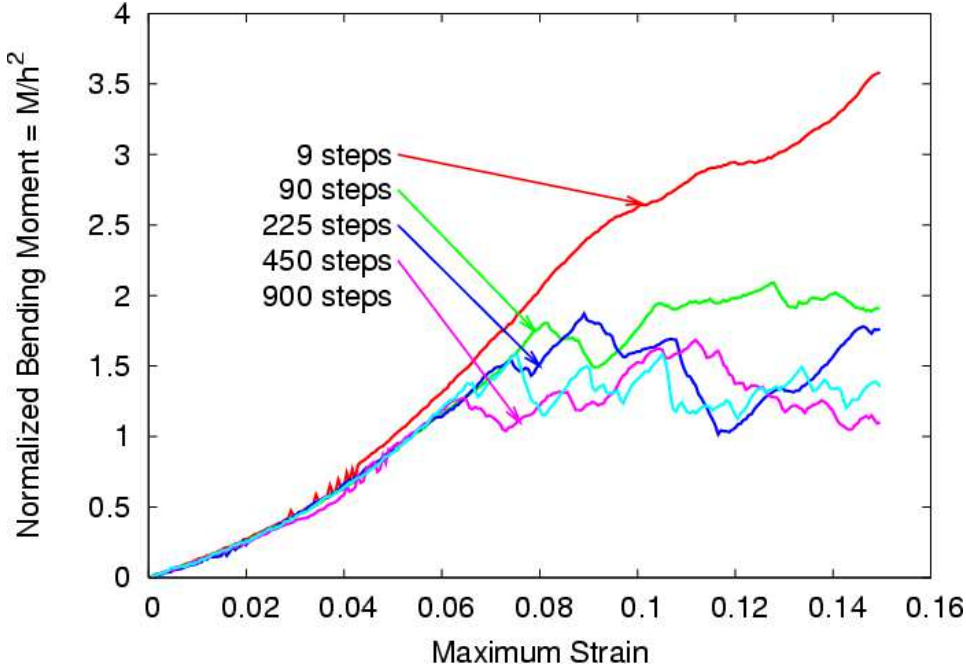}
\caption{(color) Normalized bending moment vs bending strain for different strain rates. Labels indicate the number of MC steps between applied strain increments, so that one strain increment per 9 steps is the highest strain rate, while one strain increment per 900 steps is the lowest.}
\label{fig:onesize}
\end{figure}

In initial yield, we observe an unmistakeable {\it reverse size effect}; larger crystals show a higher scaled bending moment for a given strain than smaller crystals, for a given strain rate. This behavior arises because the initial plastic response is source-limited and because the strain rate is relatively high compared to typical dislocation speed. Each sample, no matter how large, initially has only two dislocation sources at the lower two corners.  For a larger sample, it takes longer for the needed dislocations to nucleate, and because of their finite speed, it takes longer for them to glide into the interior.  Even when other sources begin operating, they are all at the boundaries.  Thus larger samples are more ``dislocation starved"~\citep{greer}, and build up higher elastic stress.

Next we consider strain rate dependence in initial yield.  Fig.~\ref{fig:onesize} shows bending moment versus bending strain for a single crystal at four different strain rates, but with the same sample size.  We again observe a transition from the elastic to the plastic regime, and note that the elastic response is independent of the strain rate.  The plastic response shows a strong rate effect, with higher strain rates leading to higher stresses and thus a higher bending moment.  Plastic response is again limited both by scarcity of sources and by the finite speed of dislocation motion relative to the effective strain rate.  Since the MC method includes no real time scale, both speed and strain rate are measured with respect to MC moves per particle.

To search for a scaling relationship between observed size and rate effects, we consider Orowan's equation, ${\dot\epsilon}=\rho v b$, where ${\dot\epsilon}$ is the strain rate, $\rho$ is the dislocation density, $v$ is the dislocation speed, and $b$ is the Burgers vector. In samples of different sizes deformed at the same strain rate, $v$ and $b$ do not scale with system size, so the dislocation densities should also be equal.  Thus if we double the sample dimensions, there should be four times as many dislocations.  Since there are initially only two sources, and the dislocation nucleation rate is roughly the same, it takes four times as long to nucleate enough dislocations to carry the plastic strain in the larger sample.  So we suppose that the size effect and rate effect in initial yield would cancel one another if we drop the strain rate by a factor of 4 each time we double the system size.

\begin{figure}[t]
\centering
\includegraphics[width=5in]{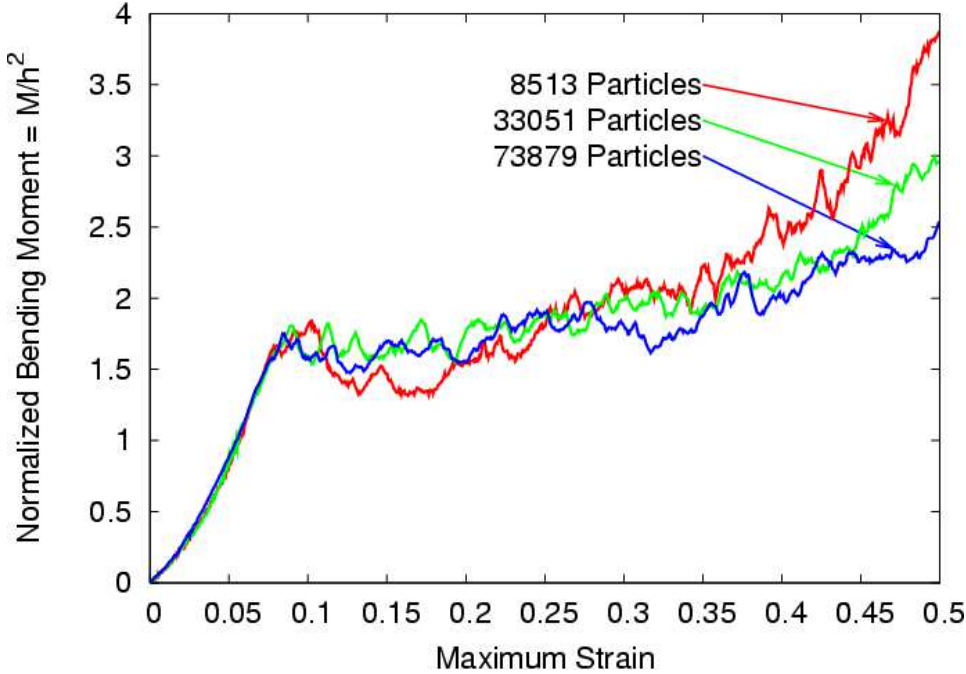}
\caption{(color) Normalized bending moment versus bending strain for three different crystal sizes, averaged over three simulation runs, with the strain rate scaled as described in the text. At low strain, the size effect is absent now that we have accounted for strain rate effects. At high strain, we observe a normal size effect of the type observed in experiment; the smallest sample has  the highest scaled bending moment and the largest sample has the lowest scaled bending moment.}
\label{fig:scaledbend}
\end{figure}

Testing this hypothesis is computationally demanding, as each factor of two in system size requires four times as many particles and four times as many MC steps, for a total factor of 16 increase in computer resources.  We carried out tests with samples ranging from approximately 90 by 90, up to 360 by 360 particles, with strain rates ranging from 25 to 400 MC steps between bend increments.  Fig.~\ref{fig:scaledbend} shows the normalized bending moment vs. strain, averaged over three simulations runs, for different sizes with the strain rate scaled as described above.  As expected, the size/rate effects appear to cancel and no size effect is observed in initial yield.  We conclude that when deformation is source-and mobility-limited, size- and rate-effects are related through the proposed scaling.

Once dislocations coalesce to form grain boundaries, a different type of size effect is observed. At high strain, Fig.~\ref{fig:scaledbend} shows a ``normal" size effect of the type observed in experiment, that is, the smallest sample shows a higher scaled bending moment and the largest sample shows a smaller scaled bending moment. Scatter in the data is due to the stochastic nature of dislocation nucleation and dynamics.  We thus conclude that two distinct size effects are observed in single crystals in two dimensions. In initial yield, there is a size effect associated with source-limited plasticity at high strain rates. Once the strain rate is accounted for by scaling, that size effect disappears. Once microstructure forms, we observe a size effect of the usual type seen in experiment.  

\subsection{Grain boundary nucleation}
We turn now to studies of dislocation coalescence to form grain boundaries.  While many researchers have performed atomistic simulations of plastic deformation on crystalline material~\citep{baskes,zuo}, most stop shortly after the onset of yield, and few have gone to high enough strains or low enough strain rates to observe formation of microstructure in a deformed single crystal.  Here we go well beyond initial yield to observe nucleation and evolution of grain boundaries.

Consider first the crystal oriented with the slip planes horizontal and at $\pm 60^\circ$, the orientation used in all the previously described simulations. Once nucleated, dislocations glide easily through the crystal and either arrest at a wall, annihilate at a free surface, dissociate into two dislocations, or combine together with another dislocation.  Total Burger's vector is conserved in both dissociation and combination, but the net Burger's vector contained in the sample increases via dislocation nucleation at the surface.  Thus a subset of dislocations in the sample are ``geometrically necessary'' to accommodate the applied bending strain.

\begin{figure}[b]
\centering
\includegraphics[width=3in]{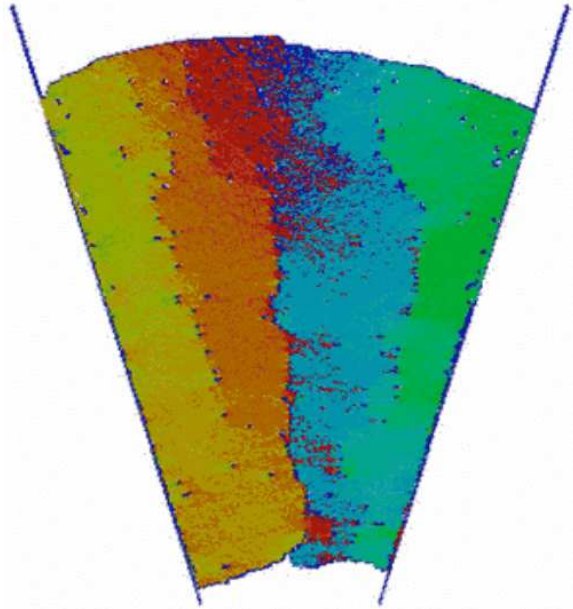}
\caption{(color) Full color image corresponding to configuration shown in Fig.~\ref{fig:bendalgorithm}(d). Particles are colored based on their underlying crystal orientation, demonstrating the formation of well-defined tilt boundaries. The angle between the two central grains is approximately 12 degrees.}
\label{fig:finalcolor}
\end{figure}

Initially, only the inclined slip planes are occupied, but both dissociation and combination give rise to dislocations on horizontal slip planes, where the stress field does not push them to glide toward a free surface.  As their numbers grow, these dislocations initially form small vertical chains, as seen in Fig.~\ref{fig:bendalgorithm}(c).  Once their density is sufficiently high, they coalesce to form well-defined tilt boundaries, as seen in Fig.~\ref{fig:bendalgorithm}(d), and in color in Fig.~\ref{fig:finalcolor}.  This dislocation patterning process has been observed in experimental studies of bending metal samples and is involved in polygonization~\citep{hirthandlothe}. To our knowledge this is the first observation of this phenomenon in an atomistic simulation.

\begin{figure}[t]
\centering
\includegraphics[width=5in]{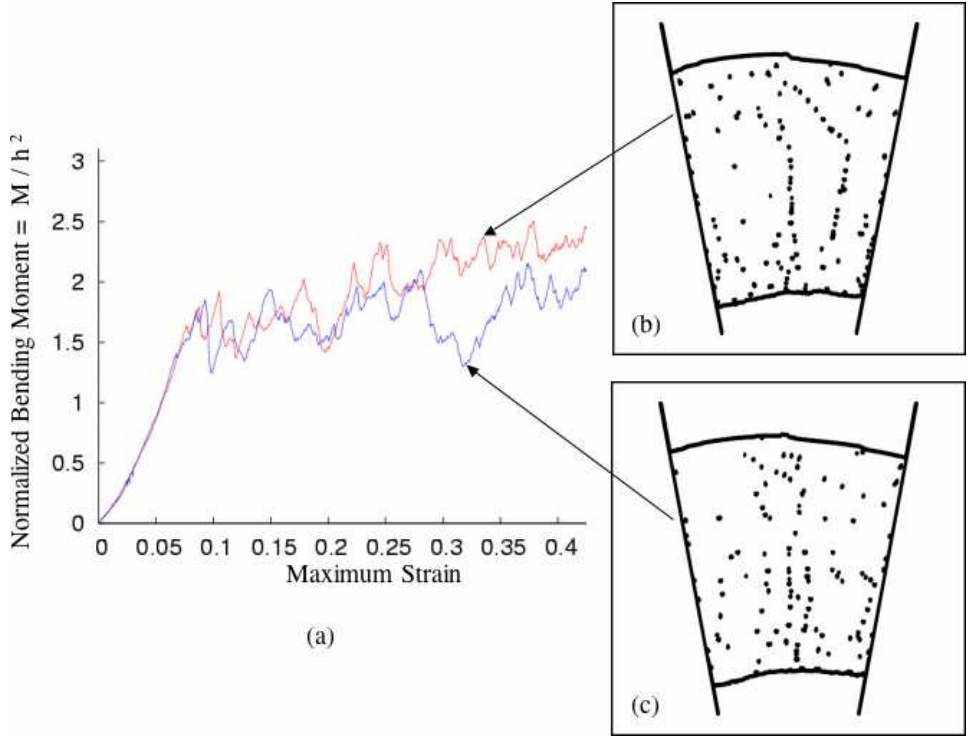}
\caption{(color) (a) Bending moment vs. strain for 73,879 particle system. The two curves represent separate simulation runs performed at the same strain rate. The configurations shown in (b) and (c) correspond to the indicated points on the curves. The higher bending moment for (b) is a result of hardening associated with dislocation microstructure.}
\label{fig:bmcompare}
\end{figure}

As mentioned earlier, each curve of bending moment vs. strain in Fig.~\ref{fig:scaledbend} is the average over three simulations.  Fig.~\ref{fig:bmcompare}(a) shows the bending moment for the 73,879 particle system as a function of strain for two separate simulations, run at the same strain rate. The microstructure in Fig.~\ref{fig:bmcompare}(b) corresponds to a higher moment than that in Fig.~\ref{fig:bmcompare}(c), as the dislocations have formed well-defined grain boundaries, leading to work hardening. The geometrically necessary dislocations (GND) that form the grain boundary are not only trapped, but also inhibit the motion of statistically stored dislocations (SSD).  The distinction between GND and SSD is a foundation of higher order strain gradient plasticity theories, and in future work we will use atomistic simulation data to explore in more detail the validity of assumptions underlying these theories.

\begin{figure}[t]
\centering
\includegraphics[width=5in]{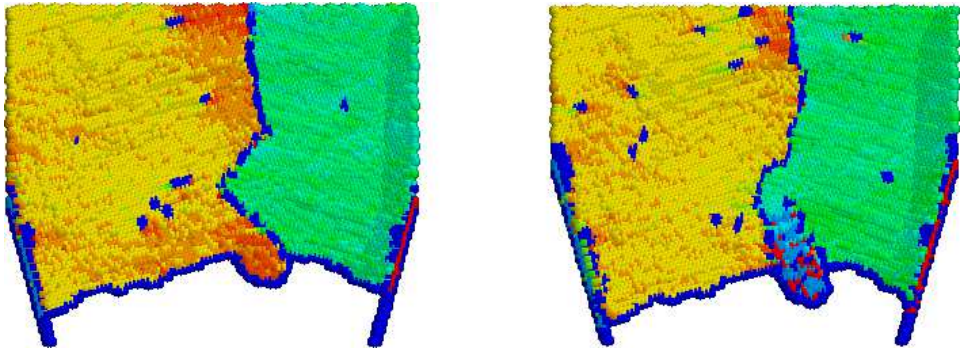}
\caption{(color) Instability in grain boundary tension causes it to fold over on itself, creating a hillock}
\label{fig:hillock}
\end{figure}

An instability in grain boundary evolution was observed and gave rise to an interesting and unexpected microstructure. When a tilt boundary reaches a high enough offset angle, it appears that absorption of further edge dislocations causes an instability in which the boundary buckles out of its vertical orientation.  We speculate that the boundary's line tension becomes increasingly anisotropic as the offset angle increases, so that the boundary eventually buckles into angled facets. In one simulation, the buckled grain boundary folds to create a new crystal grain which then protrudes from the compressive free surface, as shown in Fig.~\ref{fig:hillock} and Movie~2~\footnote{see previous notes for movie locations}.  We conjecture that this observation suggests a new mechanism for hillock formation such as that seen on a compressed Al surface ~\citep{smith}.

\subsection{Crystal orientation effects}
\begin{figure}[t]
\centering
\includegraphics[width=4in]{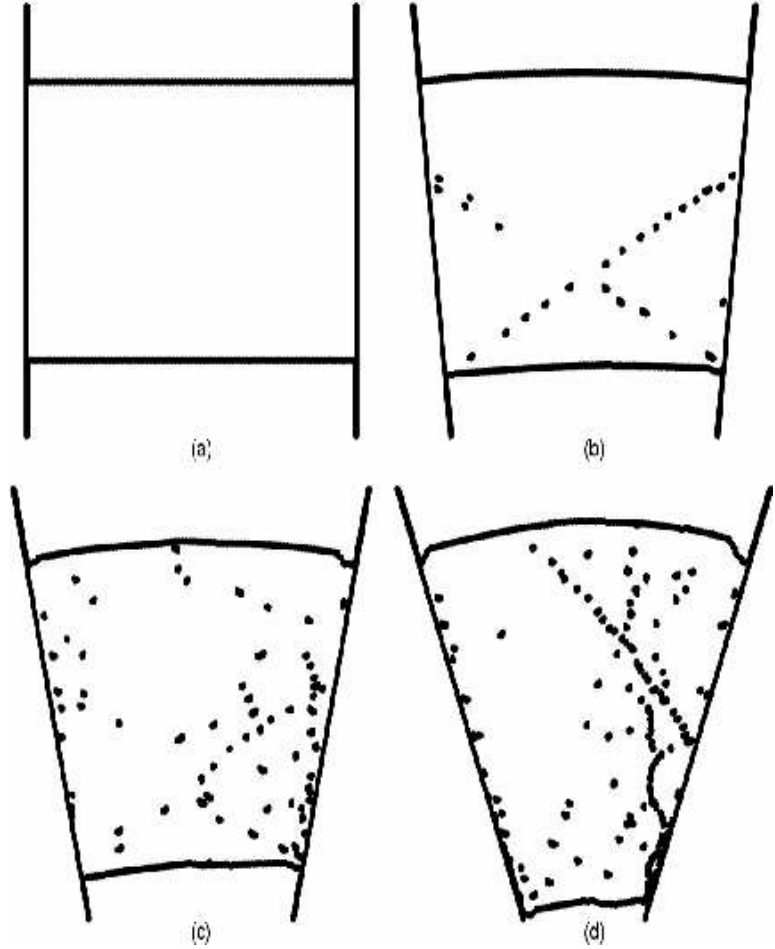}
\caption{Pattern formation changes when sample is rotated $90^\circ$ relative to that shown in Fig.~\ref{fig:bendalgorithm}.  Configuration pictured at strains of (a) 0.0, (b) 0.15, (c) 0.30, and (d) 0.50}
\label{fig:orienteffect}
\end{figure}

We performed one simulation with the crystal rotated $90^\circ$ to test the importance of lattice orientation, and found that pattern formation is both different and far less pronounced.  Dislocations emitted from opposite corners have a mutually repulsive interaction and do not combine.  At very high strains, dislocation-rich boundaries are formed, but the final structure is asymmetric with diagonal boundary formation (Fig.~\ref{fig:orienteffect}). This structure is similar to those recently observed by ~\citet{shehadeh} in multiscale simulations of dislocation interaction with shock waves at very high strain rates.

\section{Discussion}
We were initially quite surprised to observe a reverse size effect in the initial yield of single crystals, considering that every experiment we had seen in the literature pointed to a normal size effect, where smaller samples are stronger. Even considering that we were working with a simple two dimensional system, it seemed unlikely that this would alone explain the qualitative disagreement between simulation and experiment.  To make sense of this finding, we realized that the reverse size effect was produced by a unique combination of high strain rate and low source density, conditions that have not been studied in previous size effect experiments.

The reverse size effect we observe is accounted for with a scaling law in which the strain rate is a function of the system size.  This scaling law arises because there is a limited source of dislocations, namely the lower two corners, where the wall particles intersect with those in the interior. Once dislocations are nucleated, steps created on the free surface act as sources for further dislocation nucleation. This initially led us to assume that the strain rate should be scaled to the horizontal length of the sample.  However, it is only after the system has yielded that additional sources are activated, and the yield stress is not size dependent.  The dislocation density therefore must be scaled according to size, leading to a scaling law that relates strain rate to the area of the crystal.

We must keep many caveats in mind in comparing 2-d simulations to 3-d experiments. In 2-d there are only edge dislocations--point defects that interact but cannot entangle--and no screw dislocations.  Thermal activation of climb in 2-d is equivalent to moving a whole dislocation line at once and has a much higher energy barrier than in 3-d. Similar issues also affect 2-d mesoscale simulations ~\citep{groma, deshpande03, needleman}.

Our observation of dislocation coalescence into grain boundaries raises the question of whether such patterning is driven by kinetics alone.  It is possible that such microstructural patterns arise as equilibrium phases whose spacing depends only on dislocation density and temperature. One may consider a possible analogy with the twist grain boundary phase in a smectic liquid crystal, where grain boundary spacing is an equilibrium quantity rather than one determined by kinetics~\citep{chaiken}.

Our observation that ``regular" size effects set in only after a microstructure is formed is in qualitative agreement with the assumptions underlying strain gradient theory~\citep{hutch, gurtin, gao}.  Comparing our results to the theory of ``dislocation starvation" proposed by~\citet{greer} is difficult, because in our idealized two-dimensional system, dislocations may dissociate or combine but do not multiply in the interior of the sample. However our work shows that a shortage of dislocation sources may also give rise to a kind of dislocation starvation effect.

More importantly, as mentioned earlier, our results are in qualitative agreement with the experimental work of~\citet{wu}. In that study, no size effect is observed in initial yield when bending annealed gold nanowires, similar to our finding that no size effect occurs in initial yield for bending of a single crystal. We do observe a size effect at high strain, once grain boundaries form, and it would be interesting to know if the same effect would be observed in experiments on annealed nanowires strained well beyond initial yield.

We present an atomistic Monte Carlo simulation of a Lennard-Jones single crystal under applied bending in two dimensions. We observe an apparent reverse size effect in initial yield of a single crystal and argue that it arises due to a shortage of dislocation sources and high effective strain rate. The reverse size effect in initial yield disappears if the data are scaled to remove strain rate effects, such that the strain rate is reduced by a factor of four when the system size is doubled. We observe coalescence of dislocations into grain boundaries, after which a normal (smaller = stronger) size effect is observed. When a grain boundary reaches a high enough offset angle it may undergo a faceting instability which produces a protruding crystal grain resembling a hillock on a compressed metal surface. We also demonstrate that dislocation pattern formation depends on crystal orientation. Finally we speculate that the spacing of nucleated grain boundaries may not depend only on competing kinetic processes, as studied e.g. in reaction-diffusion models and DD simulations. Instead, we suggest that microstructural patterns could arise, at least in two-dimensions, as equilibrium phases whose spacing depends on dislocation density and temperature.

The authors are grateful to Drs. Lyle Levine, Robb Thomson, and Tze-jer Chuang for stimulating discussions, and to NIST-CTCMS for computer resources.  Financial support was provided by NSF DMR-0116090, NIST-CTCMS, and Catholic University.

\end{document}